\documentclass[10pt, aps,prc,twocolumn,superscriptaddress,preprintnumbers,
amsmath, 
floatfix,
longbibliography,
nofootinbib
]{revtex4-1}
\usepackage[T1]{fontenc}
\usepackage[utf8x]{inputenc} 
\usepackage{adjustbox}          
\usepackage[caption=false]{subfig}
\usepackage{url}
\usepackage{color}
\usepackage{float}
\usepackage[pdftex,colorlinks=true, linkcolor = blue, citecolor=blue,urlcolor=blue, bookmarksnumbered=true, bookmarksopen=true]{hyperref}
\usepackage{longtable}
\usepackage{amsfonts}
\usepackage{wrapfig,bm,bbm}
\newcommand{\beq}{\begin{equation}}
\newcommand{\eeq}{\end{equation}}
\newcommand{\bea}{\begin{eqnarray}}
\newcommand{\eea}{\end{eqnarray}}

\begin{document}

\title{Pre-equilibrium Neutron Emission in Fission or Fragmentation}
  
\author{Aurel Bulgac}%
\email{bulgac@uw.edu}%
\affiliation{Department of Physics,%
  University of Washington, Seattle, Washington 98195--1560, USA}
  
\date{\today}

\begin{abstract}

Fissioning nuclei and fission fragments, nuclear fragments emerging 
from energetic collisions, or nuclei probed with various
external fields  can emit one or more 
pre-equilibrium neutrons, protons, and potentially other heavier 
nuclear fragments.
%
I describe a formalism which can be used to evaluate the  
pre-equilibrium neutron emission 
probabilities and the excitation energies of the remnant fragments.
\end{abstract}

\preprint{NT@UW-20-02}

\maketitle

\section{Introduction} 

\begin{figure}[b]
\includegraphics[width=1\columnwidth]{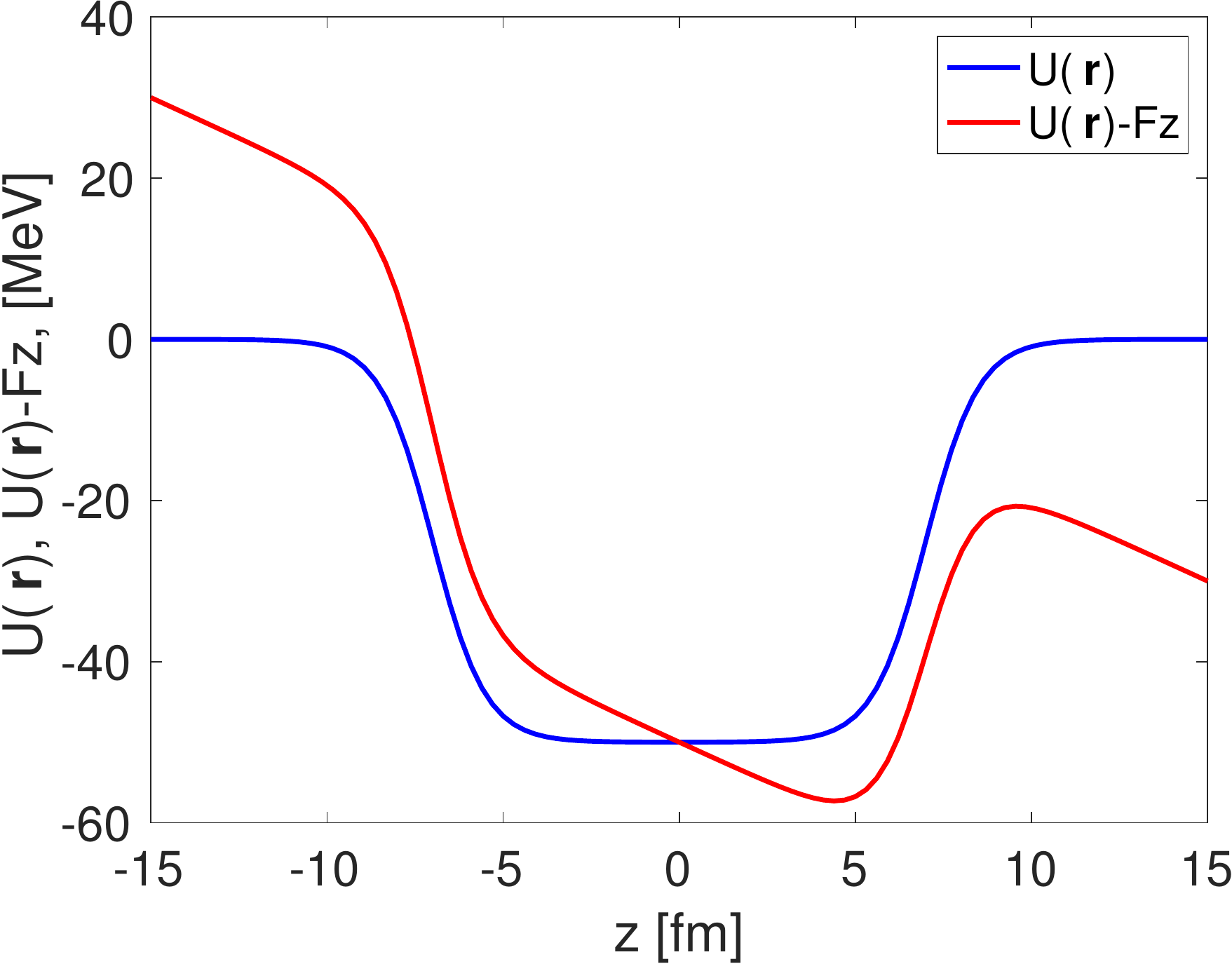}
\caption{ \label{fig:U} (Color online) The profile of the potential
experienced by nucleons at rest (black line) and of an uniformly accelerated 
one (red line) with acceleration $a=F/m$ in the $z$-direction.}
\end{figure}

Instances when particles are emitted or knocked-out of a quantum system 
after probing those systems are ubiquitous. In the Auger-Meitner effect~\cite{Meitner:1922,Auger:1923} 
in atoms, when an inner shell electron is removed, the left behind hole state is filled
by an electron from a higher energy level and the energy released is used 
to eject another electron. In nuclear physics the ejection of a deeply bound proton in 
a $(e,e'p)$, $(p,2p)$ or in a relativistic Coulomb excitation reaction 
is often accompanied by the emission of an additional 
nucleon. When a nucleus undergoes either a $\beta$- or $\alpha$-decay, the 
change of the Coulomb field of the nucleus leads to electron ionization~\cite{LL3:1977}.   
In fission, at scission immediately after the neck rupture the fission
fragments are in each other's repulsive Coulomb fields and start
accelerating and the single-particle potential experienced by nucleons
changes~\cite{Carjan:2010,Carjan:2012,Rizea:2013,Carjan:2015,Capote:2016a,Carjan:2019,Carjan:2018}, 
see also Fig. \ref{fig:U}.  The reference
framework of each fission fragment is a non-inertial one and the
equilibrium of the nuclear fluid is disturbed in a similar manner to
what happens to water in an accelerated container.  The nuclear matter
accumulates at first near the edges of the fission fragments facing
each other and at the same time the protons in the fragments are
pushed towards the opposite edges.  As a result both isoscalar and
isovector vibrational modes are excited in both
fragments~\cite{Mustafa:1971,Simenel:2014}. Nucleons are partially
promoted onto unoccupied orbitals and a fraction of them onto unbound
orbitals. The nucleons in the unbound orbitals can evaporate, in a
similar fashion to the evaporative cooling method used for decades in
cold atom experiments~\cite{Hess:1986,Roijen:1988,Masuhara:1988,
Doyle:1991,Setija:1993,Davis:1995,Petrich:1995,Cornell:2002,Ketterle:2002}. 
The goal here is to estimate the number and the probability of emitting one or
more  pre-equilibrium neutrons while the fission fragments 
are Coulomb accelerated. The formalism described here, while has number 
of similarities with previous studies quoted above, it clarifies the role of 
various approximations used and it is also extendeds in a 
number of ways the range of previously not considered in literature observables.

\section{\bf Probabilities of  pre-equilibrium neutron emission}

The time evolution of the neutrons is described by a time-dependent
Slater determinant $\Phi(x_1,\ldots,x_N,t)$ within a energy density
functional approach, which is built from the time-dependent
single-particle wave functions $\phi_k(x,t)$ with $k=1,\ldots,N$, which are solutions
of the evolution time-dependent equations
\begin{align}
i\hbar\frac{\partial\phi_k}{\partial t} = h\phi_k,
\end{align}
where $h$ is the time-dependent  mean field single-particle Hamiltonian. 
 One can consider  pre-equilibrium neutrons emitted for example from a Coulomb excited
nucleus or from a fragment in a nucleus breakup, before a compound
nucleus is formed. One possible approximation is to treat the
excitation within the Random Phase Approximation (RPA), which is the
small amplitude limit of the Hartree-Fock approximation or of the
Density Functional Theory (DFT).  If the nucleus is
weakly excited then the RPA is valid and it should be in good
agreement with the full DFT approach, which is considered here.

I will concentrate at first on the case of a single nuclear fragment
and discuss the case of two fragments below, see Eqs.~\eqref{eq:nn}
and \eqref{eq:nu}.  The neutrons populate $M>N $ bound single-particle
orbitals $\psi_n(x)$ in a fragment and $M+1,\ldots,\infty$ unbound
orbitals in a final fragment moving with the velocity ${\bm v}$
\begin{align}
{\bm v}_\text{L,R}(t)= 
\frac{1}{Nm}\langle\Phi|{\hat P}_\text{L,R}\sum_{k=1}^N\hat{ {\bm p}}_k{\hat P}_\text{L,R}|\Phi\rangle, \label{eq:v}
 \end{align}
where $\hat{ {\bm p}}_k=-i\hbar{\bm \nabla}_k$ is the momentum
operator, $m$ the nucleon mass, and 
\begin{align}
&\hat{P}_{L,R}(N)= \int_{-\pi}^\pi \frac{d\eta}{2\pi} e^{i\eta(\hat{N}_{L,R}-N) },\\
&N_{L,R}= \int d\xi  \Theta(\mp z) \sum_{n=1}^A  |\phi_n(\xi)|^2 ,
\end{align}
where $\xi={\bf r},\sigma$ stands for spatial ${\bf r}=(x,y,z)$, spin $\sigma=\uparrow,\downarrow$, and isospin $\tau=n, p$ coordinates, 
and the sum is over occupied single-particle states,
is a projector onto a
specific final nuclear fragment (the one of the left or the one on the right). Note that this projector 
${\hat P}_\text{L,R}$~\cite{Bulgac:2019x,Simenel:2010a,Sekizawa:2014}  
simply defines the part of the space where one fission fragment ends up 	(basically
$\Theta (\pm z)$ identifies the part of the volume where one or 
the other fragment is after scission, and it is different from the
projector ${\hat{\cal P}}$ introduced below.
 
I introduce now two projectors onto the final single-particle states of a 
fragment, resulting after  pre-equilibrium neutrons have been emitted.  
These projectors can be thought of as de facto analyzers of the fission 
fragment structure. For example, at
scission a typically highly excited fragment with $N_f$ neutrons was formed. This fission 
fragment can emit a number of  pre-equilibrium 
neutrons $n$, after which the remaining fragment 
will eventually turn into a 
compound nucleus with $n'=N_f-n$ neutrons, from which neutrons and gammas 
can be emitted statistically. In the first approximation one can assume that $n$ is small enough and
$n'\approx N_f$, an approximation which can be improved iteratively.
I will assume for now that no protons have been 
emitted after scission, an assumption which can be easily released if necessary.
These projectors are designed to analyze the character of the single-particle content of a 
fission fragment, specifically whether the single-particle orbitals are bound ($k=1,\ldots,M$) or 
unbound or unbound ($k>M$):
\begin{align}
& {\hat {\cal P}} = \sum_{k>M} |\psi^{\bm v}_k\rangle\langle \psi^{\bm v}_k |, \quad
{\hat {\cal Q}} = \sum_{k=1}^M |\psi^{\bm v}_k\rangle\langle\psi^{\bm v}_k |, \label{eq:pq}
\end{align}
where $\psi^{\bm v}_k(x)=\exp(im{\bm v}\cdot{\bm r}/\hbar)\psi_k(x)$
with $x={\bm r}, \sigma$ are defined in the reference frame moving
with velocity ${\bm v}({\bm r},t)$~\footnote{In the case of $^{240}$Pu
induced fission the fragments carry on average about 0.7 MeV kinetic
energy per nucleon, which amounts to a wave vector $k\approx 0.2$
fm$^{-1}$.  In this case $e^{ikR}\approx 0.36+i0.93$ and
$e^{ikd}\approx -0.73+i0.68$, where $d=2R\approx 12$ fm is the average
fission fragment diameter.  Consequently the correct values of the
overlaps $\langle \phi_k|e^{i{\bm k}\cdot {\bm r}}|\psi_l\rangle$ can
be noticeable different from an approximate estimate $\langle
\phi_k|\psi_l\rangle$.}
and 
\begin{align}
{\hat {\cal P}}+{\hat {\cal Q}}=\mathbbm{1}. \label{eq:proj}
\end{align}
The single-particle wave functions $\psi_k(x)$ describe a final 
stationary nucleus or fission fragment it its ground state 
onto which one wants to project the time evolved
single-particle wave functions $\phi_k(x,t)$.

Following a line of argument similar to the formalism described in
Refs.~\cite{Bulgac:2019x,Simenel:2010a} one can show that the
probability to have $n$ unbound neutrons is given by
\begin{align}
&P(n) = \int_{-\pi}^\pi\frac{d\eta }{2\pi}\left  \langle \Phi \left | \exp\left [i\eta( {\hat {\cal P}}-n)\right ] \right |\Phi\right \rangle\\
&= \int_{-\pi}^\pi\frac{d\eta }{2\pi} e^{-i\eta n} \det [ \delta_{kl} +( e^{i\eta}-1) \langle\phi_k| {\hat {\cal P}} |\phi_l\rangle ],\label{eq:p1}
\end{align}
and the probability that the rest of the $n'=N-n$ neutrons will be in
the $M$ bound states is
\begin{align}
\!\!\!\!\! Q(n')=  \int_{-\pi}^\pi\frac{d\eta }{2\pi} e^{-i\eta n'} 
\det [ \delta_{kl} +( e^{i\eta}-1) \langle\phi_k| {\hat {\cal Q}} |\phi_l\rangle ],\label{eq:q1}
\end{align}
and where 
\begin{align}
\langle\phi_n| {\hat {\cal P}} |\phi_m\rangle+\langle\phi_n| {\hat {\cal Q}} |\phi_m\rangle=\delta_{nm}. 
\end{align}
These formulas assume that the fission fragments were followed in time
sufficiently far enough that their accelerations at times greater than
$t$ would lead to only negligible further excitations of the nucleons
into unbound orbitals and hopefully also the one-body 
mechanism ceased to be effective~\cite{Blocki:1978,Bulgac:2019b,Bulgac:2020}.  
Since a Slater determinant is invariant under
a unitary transformation among single-particle orbitals one can always
diagonalize simultaneously the two overlap matrices $\langle\phi_n|
{\hat {\cal P}} |\phi_m\rangle$ and $\langle\phi_n| {\hat {\cal Q}}
|\phi_m\rangle$ and obtain for the probabilities $P(n)$ and $Q(n)$
simpler formulas
\begin{align}
&\langle\phi_k| {\hat {\cal P}} |\phi_l\rangle=\alpha^2_k\delta_{kl},\quad
\langle\phi_k| {\hat {\cal Q}} |\phi_l\rangle=\beta^2_k\delta_{kl}, \label{eq:ab} \\
&\alpha_k^2+\beta_k^2=1,\\
&P(n) = \int_{-\pi}^\pi\frac{d\eta }{2\pi} e^{-i\eta n} \prod_{k=1}^N [ 1+( e^{i\eta}-1) \alpha_k^2], \label{eq:P}\\
&Q(n') = \int_{-\pi}^\pi\frac{d\eta }{2\pi} e^{-i\eta n'} \prod_{k=1}^N [ 1 +( e^{i\eta}-1) \beta_k^2 ] \label{eq:Q}.
\end{align}
In the case of fission fragments the orbitals $\psi_k(x)$ with $k\le
M$ can describe the bound states in either only one or in both fission
fragments. Thus one can separate the number of neutrons emitted from
each fragment. Note that in order to calculate $P(n)$ and $Q(n)$ only
the overlaps $\langle\phi_k| {\hat {\cal Q}} |\phi_l\rangle$ between
the bound orbitals are needed.

\begin{figure}
\includegraphics[width=1\columnwidth]{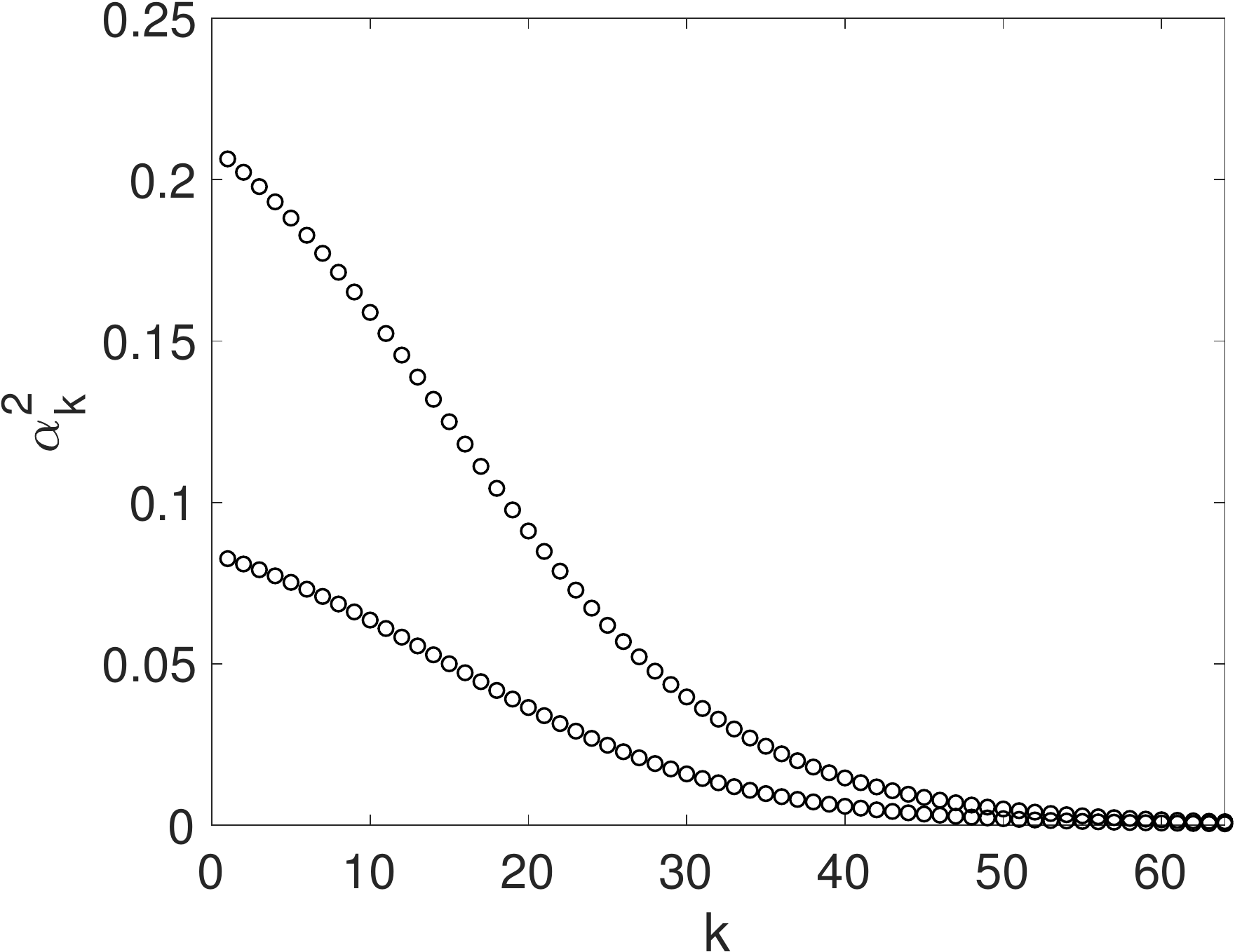}
\caption{ \label{fig:n} (Color online) Two different $\alpha_k^2$-distributions of
unoccupied overlaps $\alpha_k^2$, see Eq.~\eqref{eq:ab}, which 
differ only by their overall magnitude. $\alpha_k^2$ are fractions of the 
single-particle occupation in orbitals $\phi_k(x,t)$ lying in the continuum. One
expects that the highest lying orbitals are depleted the most and $\alpha_k^2$ are
then roughly ordered in reverse order of the instantaneous expectation value of the 
single-particle energy $\varepsilon=\langle \phi_k|h|\phi_k\rangle$, where $h$ is the 
single-particle mean field Hamiltonian. }
\end{figure}

It is useful to introduce the generating functions for the moments $\langle n^l\rangle$
and cumulates $\langle \!\langle n^l\rangle\!\rangle$~\cite{Belzig:2007,lacroix:2020}, 
which for the $P(n)$ probability distribution are          
\begin{align}
&G_\text{P}(\tau ) =   \prod_{k=1}^N [ 1+( e^{\tau}-1) \alpha_k^2]= \sum_{l=0}^\infty \frac{\tau^l}{l!}\langle n^l\rangle ,\\
&\ln G_\text{P}(\tau) = \sum_{l=0}^\infty \frac{\tau^l}{l!}\langle \!\langle n^l\rangle\!\rangle,\\
&\langle n\rangle = \sum_{k=1}^N \alpha_k^2,\quad 
\langle\!\langle n^2\rangle \!\rangle = \sum_{k=1}^N\alpha_k^2\beta_k^2, \label{eq:n2}\\
&\langle\!\langle n^3\rangle\!\rangle = \sum_{k=1}^N\alpha_k^2\beta_k^2(\beta_k^2-\alpha_k^2), \label{eq:n3}\\
& \langle\!\langle n^2\rangle \!\rangle \le \langle n\rangle, \quad 
- \langle\!\langle  n^2 \rangle\!\rangle  \le \langle\!\langle n^3 \rangle\!\rangle \le \langle\!\langle  n^2\rangle\!\rangle 
\end{align}
and one can easily to obtain symbolic expressions for higher order cumulants and similar expressions 
for the cumulants of the $Q(n')$ probability distribution.  
Two potential distributions of $\alpha_k^2$ are displayed in Fig.~\ref{fig:n}. 
As expected~\cite{Bulgac:2019x} the probabilities $P(n)$ and $Q(n')$
are correctly normalized and one can introduce the average pre-equilibrium
neutron number and its variance
\begin{align}
&\sum_{n=0}^\infty P(n) = \sum_{n'=0}^\infty Q(n')=1,\\
&\nu=\langle  n\rangle = \sum_{n=0}^\infty nP(n), \label{eq:n} \\
&\langle\!\langle n^2\rangle\!\rangle = \sum_{n=0}^\infty (n-\langle n\rangle)^2P(n).\label{eq:nvar}
\end{align}
Additionally, equivalent formulas for $P(n)$ can be derived
\begin{align}
&P(0)=\prod_{k=1}^M\beta_l^2,\\
&P(1)=P(0)\sum_{k=1}^M\frac{\alpha_k^2}{\beta_k^2},\\
&P(2)=P(0)\sum_{k>l=1}^M\frac{\alpha_k^2\alpha_l^2}{\beta_k^2\beta_l^2},\\
&P(3)=P(0)\sum_{k>l>m=1}^M\frac{\alpha_k^2\alpha_l^2\alpha_m^2}{\beta_k^2\beta_l^2\beta_m^2},
\end{align}
with similar expressions for $P(n>3)$.  

The neutron density matrix can be represented in two ways
\begin{align}
{\hat n}= \sum_{k=1}^N |\phi_k\rangle\langle \phi_k|
= \sum_{k=1}^N (  |\alpha_k\rangle +  |\beta_k\rangle)( \langle\alpha_k| +\langle\beta_k|)
\end{align}
and then show that the average number of neutrons emitted by a
fragment is
\begin{align}
&\nu=  \langle \Phi|\sum_{m=1}^\infty P(m)|\Phi\rangle = \sum_{k=1}^N\alpha_k^2.
\end{align}

\begin{figure}
\includegraphics[width=1\columnwidth]{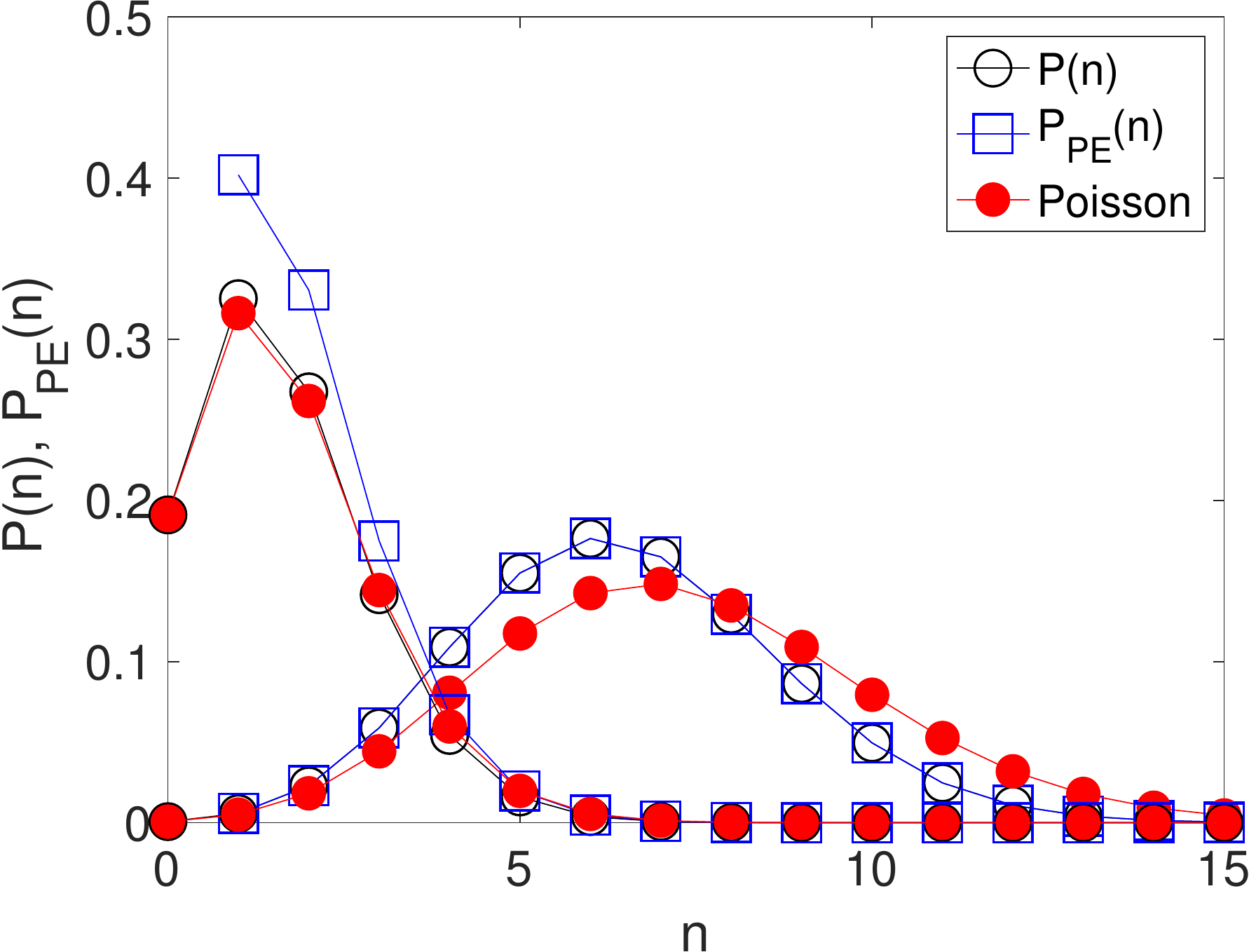}
\caption{ \label{fig:P} This figure illustrates the corresponding probabilities 
$P(n)$~\eqref{eq:P} (black circles) and $P_\text{PE}(n)$~\eqref{eq:Pc}
(blue squares) extracted using the overlap distributions from
Fig. \ref{fig:n}, and the Poisson distribution defined in Eq.~\eqref{eq:Poisson}. 
The $\alpha_k^2$-distribution with overall smaller
magnitude leads to a peak in $P(n)$ or $P_\text{PE}(n)$ with $n
\approx 1$, while the larger overlap $\alpha_k^2$-distribution have a
maximum for $n\approx 6$.  The corresponding values for
$\nu$~\eqref{eq:n} and $\nu_\text{PE}$~\eqref{eq:ncor} for these two
distributions are $\nu=$ 1.61 and 6.44, $\nu_\text{PE}=$ 1.99 and
6.44, and  the variances $\langle\!\langle n^2\rangle\!\rangle$
are 1.52 and 5.06 respectively. For the Poisson distribution the mean rate 
$\lambda =  -\ln P(0) = 1.66$ and 7.28 in these two cases and the condition
$\lambda=  \langle n\rangle =\langle\!\langle n^2 \rangle\!\rangle$ is only 
approximately fulfilled in the case of these two $\alpha_k^2$-distributions.}
\end{figure}

The knowledge of the average number of particle emitted only could
incorrectly characterize the evaporation or the decay process if $\nu$ is large. One can
envision a situation when $P(0)=1-\epsilon$ and $\sum_{n=1}^\infty
P(n)=\epsilon \ll 1$, and $P(n)$ has a weak in intensity peak at a
large $n=n_\text{max}$ value, which can be either narrow or wide. In
such a case $\nu=\langle n\rangle $ could be for example ${\cal O}(1)$
or even much smaller, even though the nucleus can emit in reality sizable
neutron clusters~\cite{Marques:2002}, but with a very low probability.
This can happen if the emitted particles can form a relatively tightly
bound cluster or a range of clusters, which are emitted with a very
low probability, a situation typical in spontaneous fission, alpha-decay or
cluster radioactivity~\cite{Rose:1984,Poenaru:1980}.  Cluster
radioactivity could be described adequately by a proper choice of the
wave functions $\psi_k(x)$.  For example, if one were to determine the
probability to form a particular type of cluster, one can chose a
density profile describing two adjacent nuclei, one with the shape of
the daughter and the other with shape of the emitted cluster. Using
the density constrained method proposed by \textcite{Cusson:1985} one
can then construct a set of single-particle wave functions $\psi_k(x)$
corresponding to such a combined density profile and define the
projector ${\hat{\cal P}}$ to select the clusters and determine their
formation probability. 

Thus the evaluation of  the entire probability distribution
$P(n)$ and not only $\nu$ can be very informative, 
in order to correctly characterize the
decay or evaporative process, see Fig. \ref{fig:P}.  If the
probability to emit no particles $P(0)$ is not small, there will be either a
weak or no correlation between $\nu$ and the value $n_\text{max}$,
where $P(n)$ is peaked and a small value of $\nu$ would merely point
to a small probability to emit many nucleons, but not characterize their
actual average multiplicity. I suggest to use instead the conditional
probability for emitting particles and define a corrected average
multiplicity neutron numbers $\nu_\text{PE}$  accordingly
\begin{align}
&P_\text{PE}(n)=\frac{P(n)}{\sum_{m=1}^\infty P(m)}, \label{eq:Pc}\\
&\nu_\text{PE}=\sum_{n=1}^\infty nP_\text{PE}(n), \quad \sum_{n=1}^\infty P_\text{PE}(n)=1.\label{eq:ncor}
\end{align}
Here $\sum_{m=1}^\infty P(m)$ is the probability that at least one
particle is emitted.  As an illustration let us consider 
the simple $P(n)$ distribution with $n_0\gg 1$
\begin{align}
&P(n) = (1-\epsilon)\delta_{n,0}+\epsilon\delta_{n,n_0},\\
&\nu = \epsilon n_0\ll n_0 ,\quad \text{but} \quad  \nu_\text{PE}=n_0.
\end{align}

A Poisson probability distribution, when the event rate is constant in time
and the events are independent, can be considered as well, and it is defined as
\begin{align}
P_\text{Poisson}(n) = e^{-\lambda}\frac{\lambda^n}{n!}, \quad
P_\text{Poisson}(0) = e^{-\lambda} \label{eq:Poisson}
\end{align}
and it is illustrated in Fig.~\ref{fig:P}. In the case of a Poisson distribution the relations
$\lambda= \langle n\rangle =\langle\!\langle n^2 \rangle\!\rangle$ are strictly satisfied.
The probability distribution $P(n)$ approaches the Poisson 
distribution when the average neutron multiplicity is $\nu\approx 1$ and smaller and 
then $\nu=\langle n\rangle \approx \langle\!\langle n^2\rangle\!\rangle$.
The Poisson limit is satisfied strictly only  in the limit $\lambda \rightarrow 0$, when 
\begin{align}
\lambda= \langle n\rangle =\lim_{\text{all} \; \alpha_k^2\rightarrow 0} \langle\!\langle n^2\rangle\!\rangle
 =\lim_{\text{all} \; \alpha_k^2\rightarrow 0} \langle\!\langle n^3\rangle\!\rangle.
 \end{align}
It is not surprising that the Poisson distribution appears quite accurate
in the mean field approximation and in the absence of fluctuations, 
see Fig. \ref{fig:P}. One should remember however that in the present analysis
the $\alpha_k^2$-distributions were considered only at a given time. Even 
in the mean field approximation there is no reason to expect that the evaporation 
rate  and the $\alpha_k^2$-distributions weakly depend on time, as the one-body 
dissipation mechanism~\cite{Blocki:1978} is effective and at work even after scission~\cite{Bulgac:2019b,Bulgac:2020}.

One can construct also the probabilities $P(n_H,n_L)$ to emit $n_H$
and $n_L$ neutrons from a heavy and a light fragments and study their
correlations.  If the projectors ${\hat{\cal Q}}_{H,L}$ on the bound
orbitals of either the heavy or light fragments $H,L$ then one define the
projectors ${\hat{\cal P}}_{H,L}=\mathbbm{1}-{\hat{\cal Q}}_{H,L}$
\begin{align}
&P(n_H,n_L) = \int_{-\pi}^\pi \frac{d\eta_H }{2\pi}  \int_{-\pi}^\pi \frac{d\eta_L }{2\pi} \nonumber \\
&\left  \langle \Phi \left | \exp \left [ i\eta_H( {\hat {\cal P}}_H-n_H) 
+i\eta_L( {\hat {\cal P}}_L-n_L)\right ]\right |\Phi\right \rangle.\label{eq:nn}
\end{align}
While constructing these projectors one should keep in mind that the
two fragments are moving with different velocities, see
Eqs.~\eqref{eq:v} and \eqref{eq:pq}.  For well separated fragments the
relations ${\hat {\cal Q}}_H{\hat {\cal Q}}_L={\hat {\cal Q}}_L{\hat
{\cal Q}}_H=0$ and ${\hat {\cal P}}_H{\hat {\cal P}}_L={\hat {\cal
P}}_L{\hat {\cal P}}_H$ are satisfied with exponential accuracy and
the final formula for $P(n_H,n_L)$ can be brought to a simple form
using the relations
\begin{align}
&\!\!\!e^{i\eta_f{\hat {\cal P}} _f}= e^{i\eta_f} - (e^{i\eta_f}-1){\hat {\cal Q}}_f, \quad \text{where} \quad f=H,L.
\end{align}
The average neutron multiplicity is given by
\begin{align}
&\nu=  \nu_H+\nu_L= N-\langle \Phi|{\hat{\cal Q}}_H|\Phi\rangle  - \langle \Phi|{\hat{\cal Q}}_L|\Phi\rangle.\label{eq:nu}
\end{align}

These formulas for neutron emission probabilities are accurate only if
the probability of emitting any protons can be neglected.  This
derivation assumes that the neutrons which populate the unbound states
are emitted before the remnant fission fragments form a compound
nuclei and are neither reabsorbed by the other fragment. A related
assumption is that once a neutron is in an unbound state it is emitted
before it has a chance to undergo any collisions in the fission
fragment and loose energy. The errors due to this last assumption can
be accounted for by using an optical potential for the neutrons in 
the unbound orbitals(as in distorted wave Born approximation). A
resonant single-particle state in the continuum is characterized by a total
width $\Gamma=\Gamma^\uparrow+\Gamma^\downarrow$, which is related to
the life-time of the state $\tau=\hbar/\Gamma$.  $\Gamma^\uparrow$ is
the escape width and its magnitude is expected to be well described within
the mean field approximation.  The spreading width $\Gamma^\downarrow$
characterizes the energy range over which the single-particle strength is
distributed~\cite{Bertsch:1983}, due to the residual interactions and can be evaluated using
an optical potential.  The probability that a particle would be
emitted, instead of loosing its energy due to in medium collisions is
proportional to the branching ratio $\Gamma^\uparrow/\Gamma$.
Therefore one can interpret the results obtained without such
corrections as upper bound estimates. An approximate way to take into
account the effect of the collisions is to replace 
\begin{align} 
\alpha_k^2\rightarrow \alpha_k^2 \times \frac{\Gamma^\uparrow}{\Gamma^\uparrow+\Gamma^\downarrow}, 
\end{align}
assuming that the spreading width $\Gamma^\downarrow$ has a weaker energy 
dependence and the branching ratio is estimated at the average energy of the
orbital $\alpha_k^2$.

Protons are also excited and can be in principle emitted as well, but
most likely only if the corresponding occupied orbitals are above the
proton Coulomb barrier. The pre-equilibrium proton emission probabilities can be
estimated in the same manner. Pre-equilibrium proton emission can be neglected
only if the corresponding probability to have all the protons in
single-particle states with energies below the Coulomb barrier is
$P(0)\approx 1$.  

Another limitation of the present approach is the
neglect of the role of fluctuations, see
Refs.~\cite{Balian:1984,Bulgac:2019a,Simenel:2011,Simenel:2018} and
references therein.

After the pre-equilibrium neutrons have been evaporated and the fission
fragments are fully accelerated, the excitation energy of the remnant
fission fragments can be used to emit neutrons and gammas from the
formed compound nucleus or fission fragments.  The number of neutrons remaining
in either the heavy or the light fragment is
\begin{align}
N_\text{H,L}=\sum_{k=1}^N\langle\phi_k | {\hat{\cal Q}}_{H,L}| \phi_k\rangle
\end{align}

One can determine the occupation probabilities $v_k^2$ in a final
fragment in its ground state in the Bardeen-Cooper-Schrieffer (BCS) approximation, under the
constraint $\sum_{k=1}^Mv_k^2=N_\text{H,L}$, and estimate the
excitation energy of such a fragment
\begin{align}
E^*_{H,L}= \sum_{k=1}^M[\langle \beta_k|h|\beta_k\rangle -v_k^2\varepsilon_k ].
\end{align}
In deriving this approximate formula I assumed that the change in the
energy is due only to the redistribution of occupation probabilities
and that the densities in the ground and excited states are basically
identical. This assumption is similar in spirit to the calculation of
the shell energy corrections due to
Strutinsky~\cite{Strutinsky:1967,BRACK:1972}.
 
All these formulas derived above implicitly assume that the average number
of neutrons remaining in the fragments after evaporation $N_{L,R}$ are
known in order to generate the single-particle wave functions
$\psi_k(x)$.  It is also implied that there exist a separation of time scales,
namely one assumes that the evaporation time - which can noticeably affected by the presence 
of the centrifugal barrier - is  shorter that the time needed to form a compound nucleus.
The wave functions $\psi_k(x)$naturally depend on
the size of the fragment, which are needed in order to evaluate
$N_{L,R}$ and $nu_{L,R}$, which satisfy the sum rule
 \begin{align}
 N=N_H+N_L+\nu_H+\nu_R. 
 \end{align}
 As typically the number of evaporated neutrons $\nu_{H,L}$ is
relatively small, one can neglect such details.  Alternatively
one can repeat the calculation once the approximate values of
$N_{L,R}$ have been determined.  It is likely that convergence can be
achieved in one or two iterations at most. If however the number of 
pre-equilibrium neutrons is relatively large one might need to repeat such a 
procedure each time after a small number neutrons are emitted.
 
Within a Hartree-Fock-Bogoliubov (HFB) framework the quasi-particle
wave functions (qpwfs) satisfy the equations
\begin{widetext}
\begin{equation} 
\label{eq:tdslda}
i\hbar \frac{\partial}{\partial t}
\left(\begin{array}{c}
u_{k\uparrow}  \\
u_{k\downarrow} \\
v_{k\uparrow} \\
v_{k\downarrow}
\end{array}\right)
=
\left(\begin{array}{cccc}
h_{\uparrow \uparrow}-\mu  & h_{\uparrow \downarrow} & 0 & \Delta \\
h_{\downarrow \uparrow} & h_{\downarrow \downarrow}-\mu & -\Delta & 0 \\
0 & -\Delta^* &  -(h^*_{\uparrow \uparrow}-\mu)  & -h^*_{\uparrow \downarrow} \\
\Delta^* & 0 & -h^*_{\downarrow \uparrow} & -(h^*_{\downarrow \downarrow} -\mu)
\end{array}\right)\left(
\begin{array}{c}
u_{k\uparrow} \\
u_{k\downarrow} \\
v_{k\uparrow} \\
v_{k\downarrow}
\end{array}\right), 
\end{equation}
\end{widetext}
where we have suppressed the spatial $\bm{r}$ and time coordinate $t$,
and $k$ labels the qpwfs $[u_{k\sigma}(\bm{r},t),
v_{k\sigma}(\bm{r},t)]$, with the z-projection of the nucleon spin
$\sigma = \uparrow, \downarrow$.  The single-particle
Hamiltonian $h_{\sigma\sigma'}(\bm{r}, t)$, and the pairing field
$\Delta(\bm{r}, t)$ are functionals of various neutron and proton
densities, which are computed from the qpwfs, and $\mu$ is the
chemical potential, see Ref.~\cite{Jin:2017} for technical
details. 

Now we have to construct the projectors onto the final (stationary) nucleus 
determined in a mean field approximation. I assume that after scission
a fragment with $N_f$ neutrons has been formed and $n$  pre-equilibrium 
neutrons are emitted and a remnant with $n'=N_f-n$ neutrons was formed. 
We will construct the ground state of the nucleus with $n'$ neutrons, and assume 
that no pre-equilibrium protons were emitted after scission. In a first approximation one can 
assume that $n$ is small enough and $n'\approx N_f$.
The quasi-particle eigenstates with $E_k>0$ (designed as occupied
quasi-particle states) are typically used to construct the nucleon
densities and the eigenstates with $E_k<0$ describe the unoccupied
quasi-particle states. For $E_k>0$ the
$v$-components and for $E_k<0$ the $u$-components of the qpwfs have a
finite norm respectively.  If $\mu<E_k<-\mu$ (as $\mu<0$ in finite
nuclei) both $v$- and $u$-components have a finite
norm~\cite{Bulgac:1980,Dobaczewski:1984,Belyaev:1987,Dobaczewski:1996}
and the spectrum is discrete.   The projectors $ {\hat {\cal P}}$ and ${\hat {\cal Q}}$
to unbound and bound  $v$-orbitals respectively are
\begin{align}
& {\hat {\cal P}} = \sum_{E_k<\mu}|\psi^{\bm v}_k\rangle\langle \psi^{\bm v}_k |, \quad
{\hat {\cal Q}} = \sum_{E_k>\mu} |\psi^{\bm v}_k\rangle\langle\psi^{\bm v}_k |, \label{eq:pq1}\\
&{\hat {\cal P}}+{\hat {\cal Q}}= \mathbbm{1},
\end{align}
where now
\begin{align}
\psi_k^{\bm v}({\bm r},\sigma) =
\begin{pmatrix}
&u_{k\uparrow}({\bm r})e^{\frac{i m{\bm v}\cdot{\bm r}}{\hbar}} \\
&u_{k\downarrow}({\bm r})e^{\frac{i m{\bm v}\cdot{\bm r}}{\hbar}}  \\
&v_{k\uparrow}({\bm r})e^{-\frac{i m{\bm v}\cdot{\bm r}}{\hbar}}  \\
&v_{k\downarrow}({\bm r})e^{-\frac{i m{\bm v}\cdot{\bm r}}{\hbar}} \end{pmatrix},
\end{align}
as under a boost the $u$- and $v$-components of the qpwfs
transform in opposite manner~\cite{Stetcu:2014,Nakatsukasa:2016}. This aspect is also
manifest in the structure of the time-dependent density functional theory (TDDFT) Eqs.~\eqref{eq:tdslda}, as the
single particle Hamiltonian changes under a boost as
$h_{\sigma,\sigma}\rightarrow h_{\sigma,\sigma}+{\bm v}\cdot{\hat{\bm
p}}$~\cite{Stetcu:2014}.

The projector ${\hat {\cal Q}}$ projects on both occupied and
unoccupied bound quasi-particle states, for which $\int d{\bm r}
|v_k({\bm r},\sigma)|^2 <\infty$ in the final nucleus or in the
fission fragment. If the sum in the definition of ${\hat{\cal Q}}$
would have been restricted to $E_k>0$, only the occupied quasiparticle
states in the ground state of the nucleus or fragment would have been
included.  In the case of a HFB framework the quasiparticle spectrum
is continuous for both occupied and unoccupied quasi-particle states
if $|E_k|>|\mu|$ and the projector ${\hat {\cal P}} $ selects only the
unbound unoccupied quasi-particle states with $E_k<\mu$, when $\int d{\bm r}
|v_k({\bm r},\sigma)|^2 \rightarrow \infty$.
 
The Eqs.~\eqref{eq:p1} and \eqref{eq:q1} read in this
case~\cite{Bulgac:2019x}
\begin{align}
&P(n)=\int_{-\pi}^{\pi} \frac{d\eta}{2\pi} e^{-i\eta n}\sqrt{ \det{\left [ \delta_{kl} +(e^{i\eta}-1) P_{kl}  \right ] }},\label{eq:P2}\\
&Q(n')=\int_{-\pi}^{\pi} \frac{d\eta}{2\pi} e^{-i\eta n'}\sqrt{ \det{\left [ \delta_{kl} +(e^{i\eta}-1) Q_{kl}\right ] }},\label{eq:q2}
\end{align}
where 
\begin{align}
P_{kl}=\langle \phi_k|{\hat{\cal P}}|\phi_l\rangle,\quad
Q_{kl}=\langle \phi_k|{\hat{\cal Q}}|\phi_l\rangle
\end{align} 
and $\phi_k(x)$ are now the 4-components  Bogoliubov quasiparticle wave
functions obtained by evolving Eqs.~\eqref{eq:tdslda}.  
After orthogonalizing $\langle \phi_k|{\hat{\cal P}}|\phi_l\rangle$ and $\langle \phi_k|{\hat{\cal
Q}}|\phi _l\rangle$ these expressions simplify
\begin{align}
&P(n)=\int_{-\pi}^{\pi} \frac{d\eta}{2\pi} e^{-i\eta n}\sqrt{ \prod_{k=1}^{2\Omega} \left [1+(e^{i\eta}-1)\alpha^2_k\right ] },\\
&Q(n')=\int_{-\pi}^{\pi} \frac{d\eta}{2\pi} e^{-i\eta n'}\sqrt{ \prod_{lk1}^{2\Omega} \left [1+(e^{i\eta}-1)\beta^2_k\right ] },
\end{align}
where 
\begin{align}
\alpha_k^2= \langle \phi_k|{\hat{\cal P}}|\phi_k\rangle, \quad 
\beta_k^2 = \langle \phi_k|{\hat{\cal Q}}|\phi_k\rangle,
\end{align}
and $2\Omega$ 
is the dimension of the Fock space.  The total number of pre-equilibrium
neutrons evaporated can be determined either from
$\nu=\sum_{n=0}^\infty nP(n)$
or as
\begin{align}
\nu=\sum_k\langle \phi_k|{\hat{\cal P}}|\phi_k\rangle.
\end{align}
If instead one uses a TDHF-BCS framework~\cite{Scamps:2012a,Scamps:2013,Simenel:2018,Sekizawa:2019} to describe the initial 
nucleus then 
\begin{align}
v_k(x,t)=v_k(t)\phi_k(x,t),\quad u_k(x,t)=u_k(t)\phi_k(x,t)
\end{align} 
$|v_k(t)|^2$ are the occupation probabilities, $\phi_k(x,t)$ are 
2-components single-particle wave functions obtained as
solutions of the TDHF equations, $\langle\phi_k|\phi_l\rangle=\delta_{kl}$, and $|v_k(t)|^2+|u_k(t)|^2=1$.

\section{Conclusions}

The formalism outlined here can be used to characterize the fate of the
quasi-particle states promoted into the continuum either in an excited nucleus 
or in an excited nuclear fragment. One can calculate
for each quasiparticle state, initially localized inside the nucleus,
an averaged transmission probability into the continuum.
These transmission probabilities lead to upper estimates of the number of the pre-equilibrium
neutrons emitted, up to corrections due to the branching ratio $\Gamma^\uparrow/\Gamma$.
The only other source of uncertainties is due to the role
of fluctuations, which is expected to lead to wider 
distributions, but it will likely not affect radically the average neutron
multiplicities~\cite{Balian:1984,Bulgac:2019a,Simenel:2011,Simenel:2018}. The role of fluctuations can 
be accounted for in a variety of ways~\cite{Grange:1983,Weidenmuller:1984,Frobrich:1998,Randrup:2011,
Ward:2017,Sierk:2017,Ishizuka:2017,Albertsson:2020,Bulgac:2019a}. The extension to emission 
of other kind of particles is straightforward. 

The formalism can be extended to project other single-particle properties, such as the  
energies of the emitted  nucleons, and/or their angular momenta in a manner discussed in 
Refs.~\cite{Bulgac:2019x,Sekizawa:2014,Avez:2013}~\footnote{In Ref.~\cite{Avez:2013} 
the bound and unbound nucleons are discriminated by their spatial distribution, 
not by their single-particle energies, as suggested here, and that can lead in principle to different results.}. 
By projecting the linear momenta of the emitted nucleons one can obtained simultaneously
the angular and the energy distributions of the emitted nucleons.

\vspace{0.5cm}

\section*{\bf Acknowledgements} 
I thank G.F. Bertsch for discussions and N. Carjan 
and I. Stetcu for urging me to think about this problem.
This work was supported by U.S. Department of Energy,
Office of Science, Grant No. DE-FG02-97ER41014 and in part by NNSA
cooperative Agreement DE-NA0003841. 


\providecommand{\selectlanguage}[1]{}
\renewcommand{\selectlanguage}[1]{}

\bibliography{latest_fission}

\end{document}